\pdfoutput=1
\documentclass[useAMS,usenatbib]{mn2e}
\usepackage{graphicx}
\usepackage{times}
\usepackage{natbib}
\usepackage{subfigure}
\usepackage{url}
\usepackage{float,lscape}
\usepackage{amsmath}
\usepackage[dvipsnames]{xcolor}
\usepackage{amssymb}
\usepackage{multirow}
\usepackage{tikz}
\usepackage[titletoc,page]{appendix}

\def\hii{\mbox{H\,{\sc ii}}}
\def\hi{\mbox{H\,{\sc i}}}
\DeclareMathOperator{\taninv}{tan^{-1}}

\title[\textrm{\hi}~Jet-like Feature Towards G351.7--1.2]{High Velocity \hi~Jet-like Feature Towards the SNR Candidate G351.7--1.2}

\author[Veena et al.]{V. S. Veena$^{1}$\thanks{E-mail: veenavs.13@iist.ac.in}, Sarita Vig$^1$, Nirupam Roy$^2$, Jayanta Roy$^3$\\
$^1$Department of Earth and Space science, Indian Institute of Space Science and Technology, Thiruvananthapuram, 695 547, India\\
$^2$Department of Physics, Indian Institute of Science, Bangalore, 560 012, India\\
$^3$National Centre for Radio Astrophysics (NCRA-TIFR), Pune, 411 007, India\\}

\begin{document}

\date{}

\pagerange{\pageref{firstpage}--\pageref{lastpage}} \pubyear{}

\maketitle

\label{firstpage}


\begin{abstract}
We present the \hi~21~cm spectral line and continuum observations of the Galactic supernova remnant candidate G351.7--1.2 using the upgraded Giant Metrewave Radio Telescope. Strong absorption features are observed towards the \hii~regions in the star forming complex associated with G351.7--1.2. Along with \hi~emission towards the outer periphery of the supernova remnant shell, we distinguish a high velocity jet-like feature in the velocity range $+40$~km~s$^{-1}$ to $+52$~km~s$^{-1}$ in \hi. This unusual and highly collimated feature, with a projected length of $\sim$7~pc and an opening angle of $14.4^\circ$, is located  towards the interior of the radio shell. This is the first report of a such a well collimated \hi~jet-like emission. The peculiar location and the detection of a $\gamma$-ray source towards the central peak of this HI jet suggests its plausible association with the supernova remnant candidate.  

\end{abstract}

\begin{keywords}
ISM: supernova remnants -- ISM: \hii~regions -- radio continuum: ISM -- radio lines: ISM -- ISM: individual: SNR~G351.7--1.2
\end{keywords}

\section{Introduction}

Supernova remnants (SNRs) form as a result of the interaction between the supernova ejecta and the ambient interstellar medium (ISM). During the violent death of a massive star as a supernova, it releases a few solar mass ejecta and $\sim10^{51}$~ergs of kinetic energy into the ISM \citep{1984A&A...133..175H}, forming a SNR. Majority of the remnants of core-collapse supernovae are expected to evolve in regions associated with dense molecular clouds owing to the short life span of their massive progenitors. The blast waves from supernova explosions have a destructive effect in their immediate surroundings as they tend to disrupt the parental molecular cloud. However, supernovae could also initiate triggered star formation in swept up clouds that are dense enough to survive the explosion \citep{{1977ApJ...217..473H},{2009MNRAS.397.1215P}}. Such SNR-cloud systems provide vital clues in understanding the complex interaction mechanisms between SNR and ambient molecular cloud.  

\par The massive star forming complex G351.7--1.2 includes two \hii~regions IRAS 17258--3637 and IRAS 17256--3631 \citep{{2014MNRAS.440.3078V},{2016MNRAS.456.2425V},{2017MNRAS.465.4219V}}. The wide band radio continuum observations towards this star forming complex using upgraded Giant Metrewave Radio Telescope \citep[uGMRT;][]{gupta2017upgraded} have confirmed the presence of large scale diffuse emission (angular extent $\sim$14$'$) with a broken shell morphology encompassing the \hii~regions. The spectral indices of the \hii~regions are consistent with thermal free-free emission whereas spectral index of the diffuse shell is found to be $-0.79$ indicating non-thermal emission. The radio shell is associated with morphologically similar H$\alpha$ emission and a Fermi LAT $\gamma$-ray source, 1FGL1729.1--3641c, is also identified towards south-west of the radio shell. Shock heated mid-infrared dust emission is seen towards the region and cold dust emission is detected only towards the west that could explain the broken-shell morphology of this object. Based on our multiwavelength analysis, we concluded that G351.7--1.2 (hereafter G351.7) is a candidate SNR that has not been reported previously \citep[][hereafter Paper I]{2019MNRAS.482.4630V}. \\

\par The \hi~21~cm observations of the candidate G351.7 (Proposal code: 34$\_$014) carried out using the uGMRT is presented in this work. We use \hi~imaging to detect the expanding shells of \hi~around the SNR candidate. 

\section{GMRT Observations and Data Reduction}

\subsection{21~cm Observation}
We have mapped the \hi~21 cm emission towards G351.7--1.2 using the Giant Metrewave Radio Telescope, India \citep{1991CuSc...60...95S}. The observed field was centred at the candidate SNR G351.7--1.2 ($\alpha_{J2000}$: $17^h29^m19.0^s$, $\delta_{J2000}$: $-36^\circ37\arcmin10.0\arcsec$). As our target is a southern source and is available $\sim7$~hrs per day above the elevation limit of GMRT, the target was observed for 6 days (May 10-11, June 3-5, September 21, 2018) resulting in an on-source integration time of 27~hrs. The rest frequency of \hi~21 cm line is 1420.4057~MHz. The observing frequency was estimated keeping in view the LSR velocity of $\sim-12$~km~s$^{-1}$ of this region \citep{2017MNRAS.465.4219V} as well as the motions of the Sun and the Earth. The bandwidth of the observations was 4~MHz, divided into 512 channels. This corresponds to a spectral resolution of 8~kHz (velocity resolution of 1.7~km~s$^{-1}$).  The radio sources 3C286 and 3C48 were used as the primary flux calibrators whereas 1626-298 was used as phase calibrator. We have also used 3C286 for bandpass calibrations.

\begin{figure}
\centering
\includegraphics[scale=0.33]{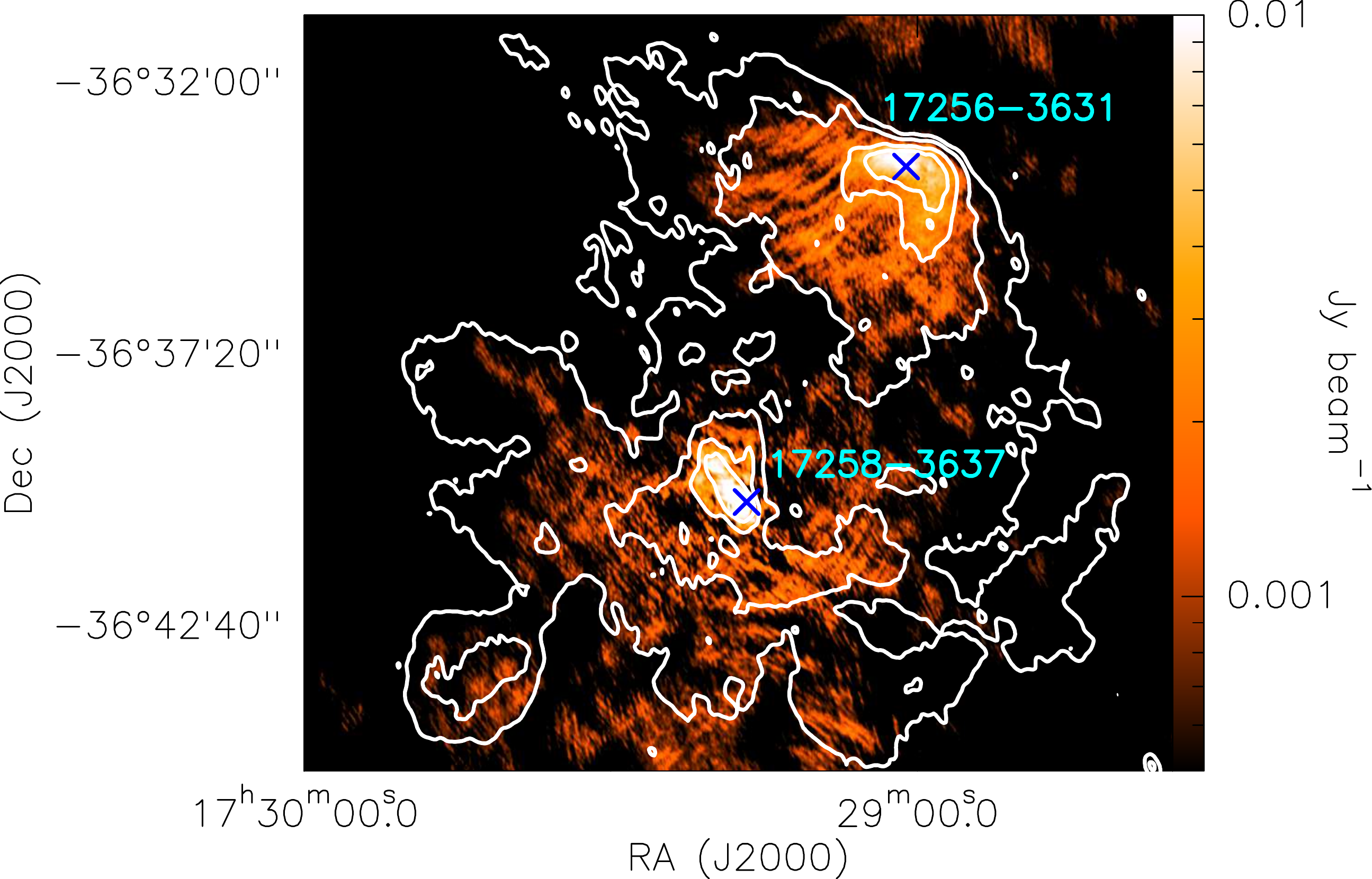} \hspace*{0.5cm} \includegraphics[scale=0.35]{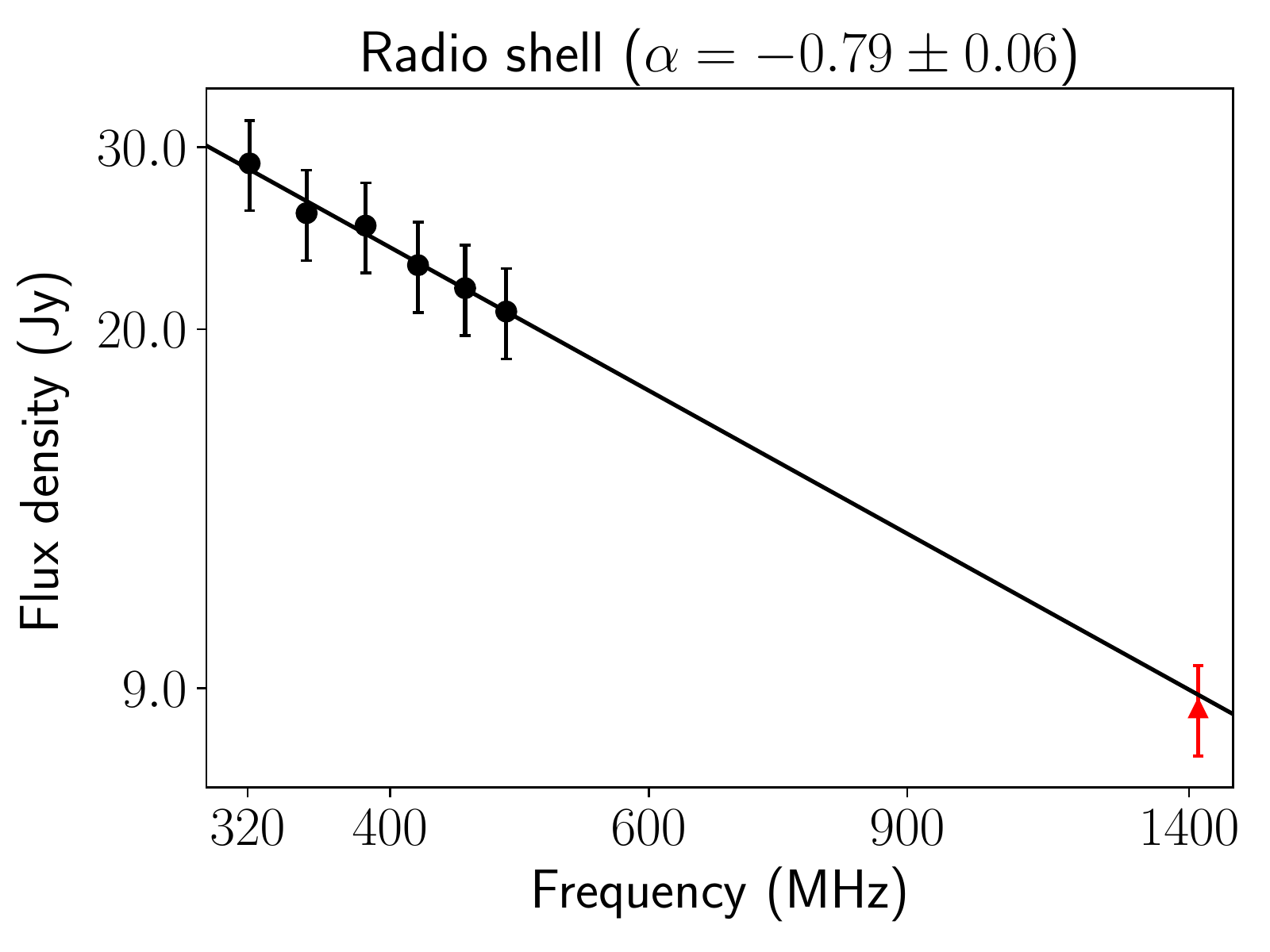}
\caption{(Top) 21~cm continuum image of G351.7 overlaid with 321~MHz radio contours. The contour levels are 16, 50, 150 and 240~mJy\,beam$^{-1}$. Peak positions of the two \hii~regions IRAS~17256--3631 and IRAS~17258--3637 are marked as $\times$. The beam is $9.6\arcsec\times7.4\arcsec$. (Bottom) Spectra of the radio shell using fluxes from 321 to 480 MHz (Paper I). Triangle shows the flux at 1420 MHz from the current work.}
\label{21continuum}
\end{figure}

\par The data reduction was carried out using the NRAO Astronomical Image Processing System ($\tt{AIPS}$). Each of the six data sets were carefully checked for radio frequency interference and corrupted data due to non-working antennas, bad baselines etc. The flagging (removal) of such data points were carried out and the calibrated data sets were then cleaned and deconvolved to produce a continuum map. We subtracted the continuum (from line free channels) to create spectral line UV data. The six data sets were combined and a spectral cube was generated. In order to have better sensitivity to diffuse extended emission, we have created a low resolution map by considering $\tt{UVTAPER}$ of 4~k$\lambda$ in $\tt{IMAGR}$. A rescaling factor (T$_\textrm{gal}$+T$_\textrm{sys}$)/T$_\textrm{sys}$ is used to correct for the contribution of Galactic plane emission to the antenna system temperature. T$_\textrm{sys}$ corresponds to the system temperature associated with the flux calibrators that is located away from the Galactic plane. T$_\textrm{gal}$ is computed by extrapolating the sky temperature value at 408~MHz \citep{1982A&AS...47....1H} to 1420 MHz and assuming a spectral index of $-2.6$ \citep{{1999A&AS..137....7R},{2011A&A...525A.138G}} for the Galactic plane emission. The estimated rescaling factor of 1.2 is applied and the image is then primary beam corrected. The final resolution of the spectral cube is $48.6\arcsec\times32.7\arcsec$, while that of continuum map is $9.6\arcsec\times7.4\arcsec$.

\subsection{Pulsar observation}

To search for possible pulsars associated with the supernova remnant candidate G351.7, we have carried out DDT pulsar observations using uGMRT (Proposal code: DDTC032) on 09$^{th}$ of December, 2018. We used GMRT band 4 (550--750~MHz) in the incoherent array mode with a bandwidth of 200 MHz covered through 4096 channels. The radio signals from the observing band were converted into baseband signals and are then delay correlated and Fourier transformed to obtain the spectral information. 3C286 was used as the primary flux calibrator whereas 1626-298 was used as secondary phase calibrator for imaging observations. The total on-source integration time was 170 minutes. We recorded incoherent array filter-bank data with a time resolution of 163.84~$\mu$s. The uGMRT band 4 incoherent array beam of size $40\arcmin-50\arcmin$ completely covers the positional uncertainty of the $\gamma$-ray source (Fermi 95$\%$ confidence region of 4.5$\arcmin$). Considering the distance to SNR as 2~kpc, the line-of-sight dispersion measure (DM) is 100 pc cm$^{-3}$ \citep{2002astro.ph..7156C}. We searched for pulsation up to DM of 500 pc cm$^{-3}$. 

\begin{figure}
\centering
\includegraphics[scale=0.34]{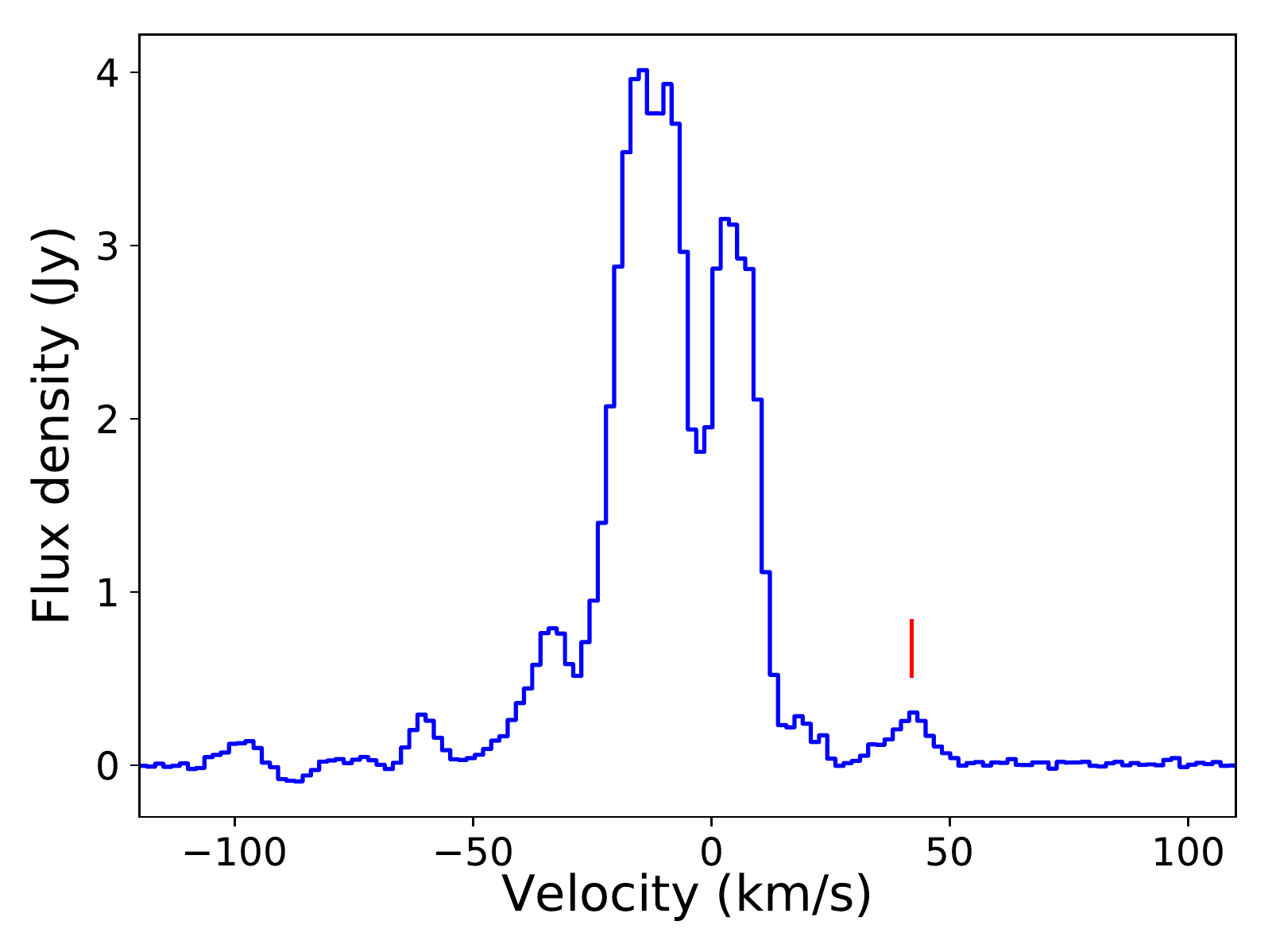} \quad \hspace*{0.5cm} \includegraphics[scale=0.34]{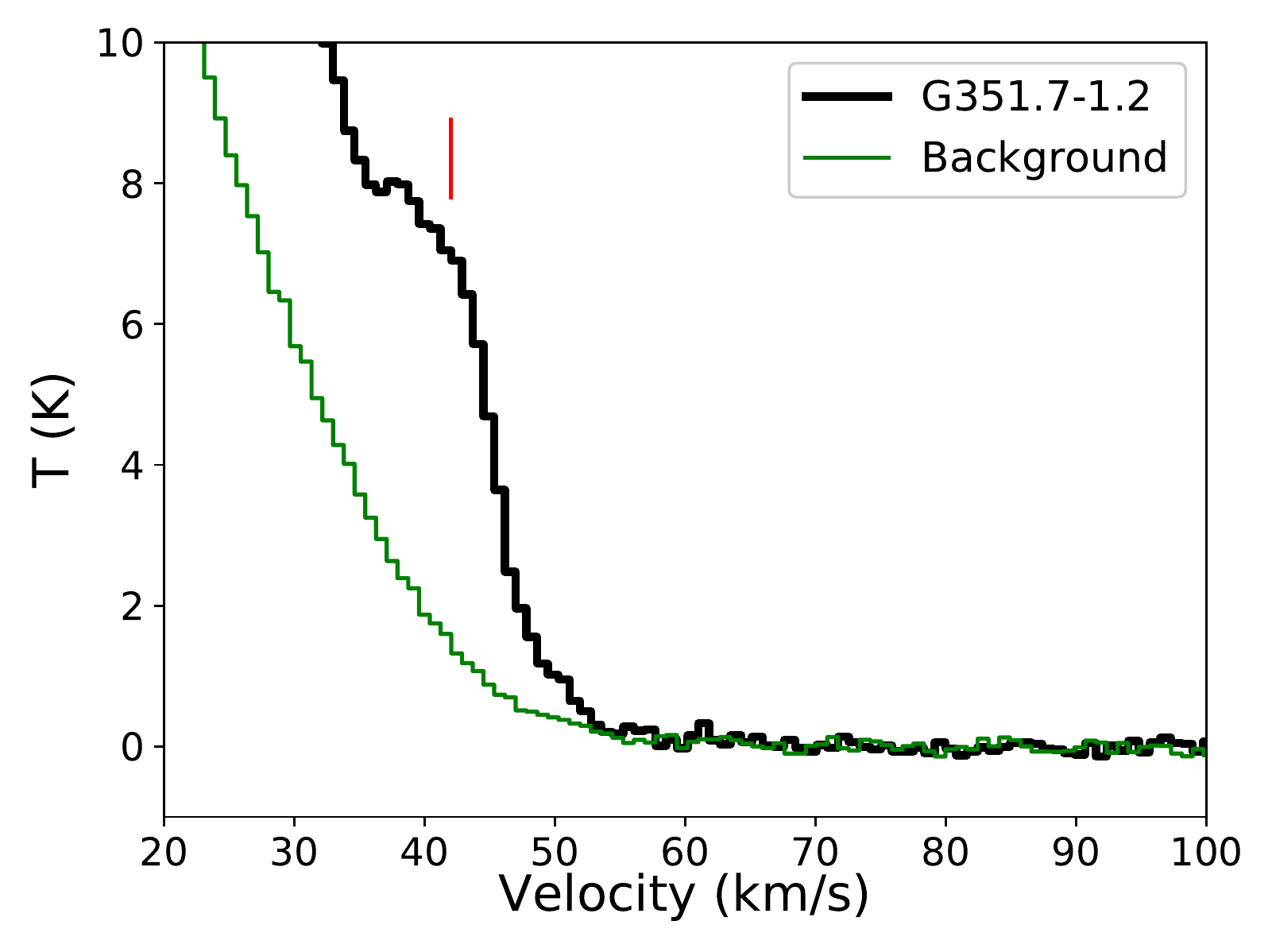}
\caption{(Top) Spectra of the \hi~21 cm emission towards the SNR candidate G351.7. (Bottom) GASS \hi~spectrum of the high velocity feature towards G351.7--1.2 over-plotted with the background spectrum.  Solid line indicates the high velocity feature. }
\label{spec}
\end{figure}

\section{Results and Discussion}


\subsection{Radio Continuum Emission}

The continuum image of G351.7 at 1420.5~MHz is shown in Fig.~\ref{21continuum} (Top). The most prominent features in the figure are the two \hii~regions IRAS 17256--3631 and IRAS 17258--3637. The diffuse shell structure of the SNR candidate identified at low frequencies (see Fig.~2 of Paper I) is not evident in this image. Faint radio continuum emission from part of the radio shell is observed towards south and south-east of IRAS 17258--3637. The angular extent of this diffuse emission is $\sim$7$\arcmin$. The minimum UV distance corresponding to the target at 21~cm is 250~$\lambda$. This corresponds to an angular size of 13.7$\arcmin$. The size of the radio shell at 321 MHz is $\sim$15$\arcmin$ and even though the size is comparable to the largest angular structure, there could be missing flux. Fig.~\ref{21continuum} (Bottom) shows the radio spectra of the non-thermal shell from Paper I over-plotted with the flux density at 1420 MHz. In Paper I, we estimated the spectral index of the diffuse shell (excluding thermal emission from the HII regions) as $-0.79$. The flux density at 1420 MHz is broadly  consistent with the non-thermal spectral index obtained from lower frequencies, as evident from the figure. However, if there is missing flux at 21 cm, this could contribute to  thermal emission towards the shell. With the present data, we cannot not estimate the contribution of thermal and non-thermal emission from the SNR shell.

\subsection{\hi~Emission}

The spectrum of \hi~21 cm emission towards G351.7 is shown in Fig.~\ref{spec} (Top). G351.7 is located in the third Galactic quadrant where most of the \hi~emission is expected to be confined to the velocity range $-200$~km~s$^{-1}$ to $+30$~km~s$^{-1}$ according to the Galactic rotation model. However, we notice an additional peak at the LSR velocity $\sim+42$~km~s$^{-1}$ that cannot be explained using the rotation curve model of the Milky Way \citep[V$_{max}\sim$30~km~s$^{-1}$ at $l=351.7^\circ$;][]{1989ApJ...342..272F}. This is also evident in the Parkes Galactic All-Sky Survey spectrum \citep[GASS;][]{2009ApJS..181..398M}. A comparison of the GASS spectra (angular resolution $16\arcmin$) towards the target and a background region located $\sim$20$\arcmin$ away is presented in Fig.~\ref{spec} (Bottom). The target spectrum shows an additional peak at $\sim$40~km~s$^{-1}$ towards G351.7. This is consistent with the spectrum obtained from GMRT.

\begin{figure}
\centering
\includegraphics[scale=0.33]{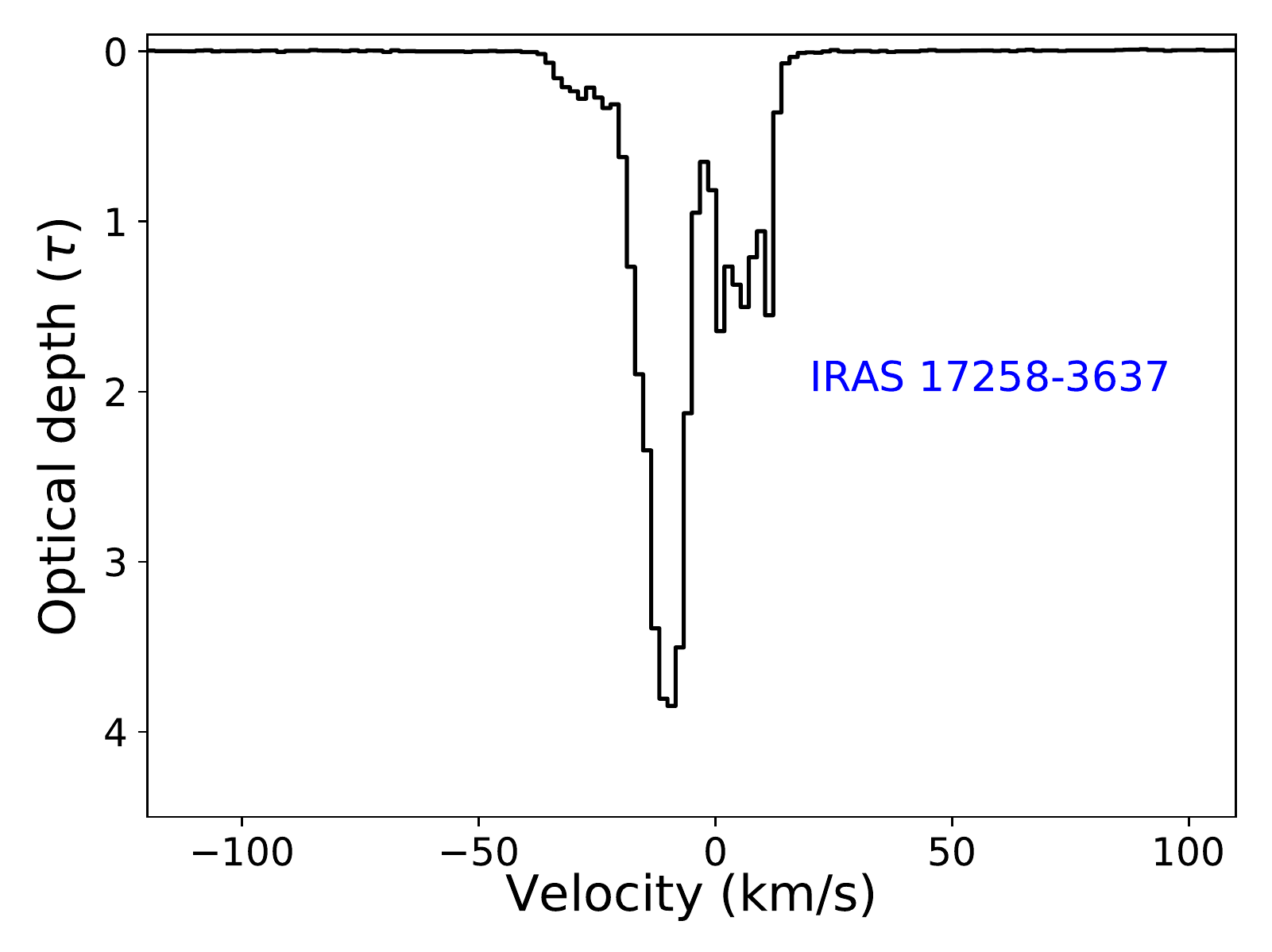} \quad \includegraphics[scale=0.33]{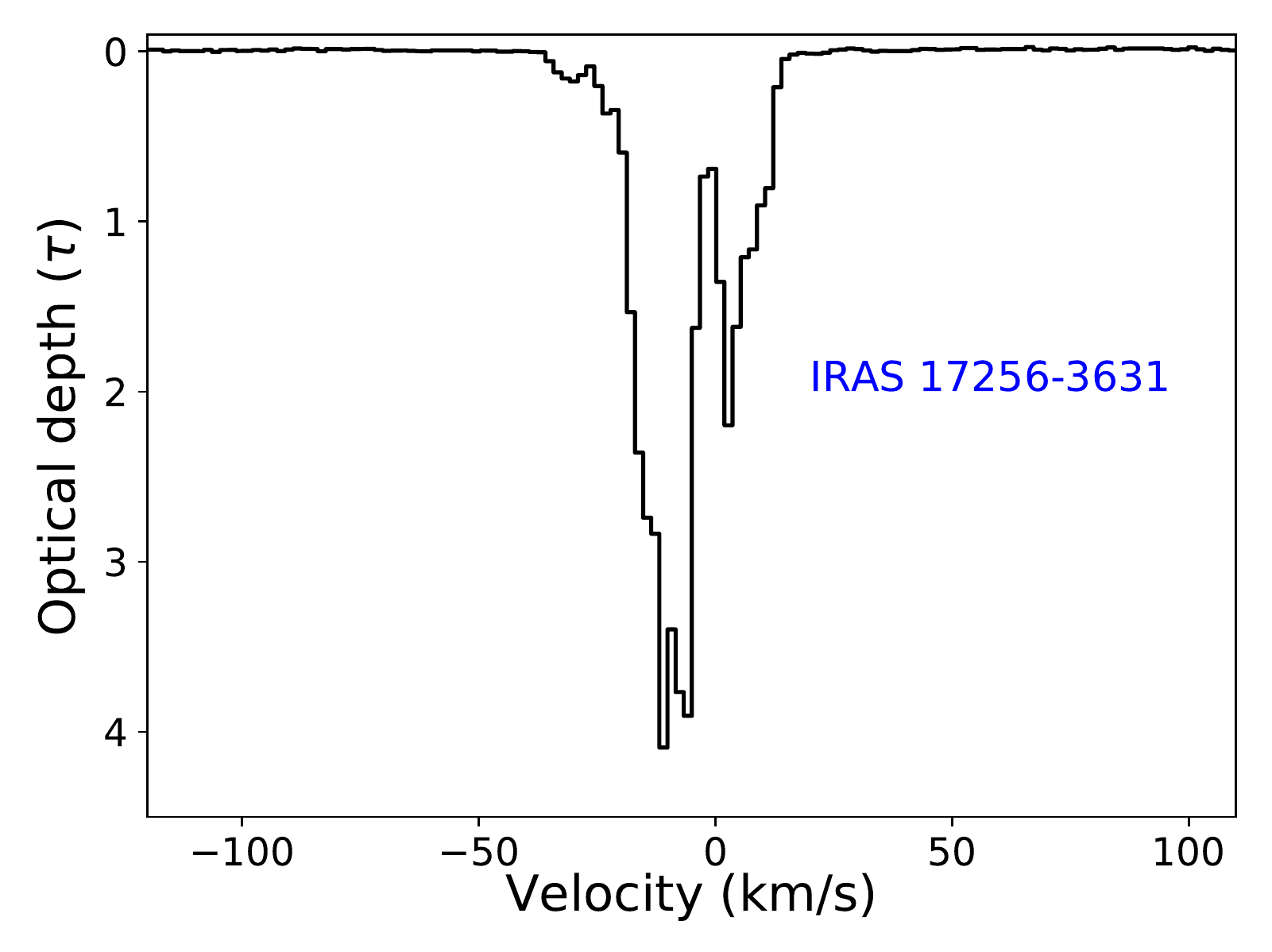}
\caption{21~cm \hi~absorption spectra towards the peak of the \hii~regions (Top) IRAS 17258--3637 (Bottom) IRAS 17256--3631.}
\label{vel_low}
\end{figure}

Fig.~\ref{vel_low} shows the spectra towards the two \hii~regions IRAS 17258--3637 and IRAS 17256--3631.  The \hi~absorption against the continuum in the \hii~regions peaks at V$\sim-12$~km~s$^{-1}$, corresponding to the LSR velocity of these \hii~regions. The result is consistent with the LSR velocity obtained from recombination line observations \citep{2017MNRAS.465.4219V}. Absorption features are also observed towards the diffuse radio shell in the similar velocity range as that of the \hii~regions.

\begin{figure*}
\centering
\includegraphics[scale=0.72]{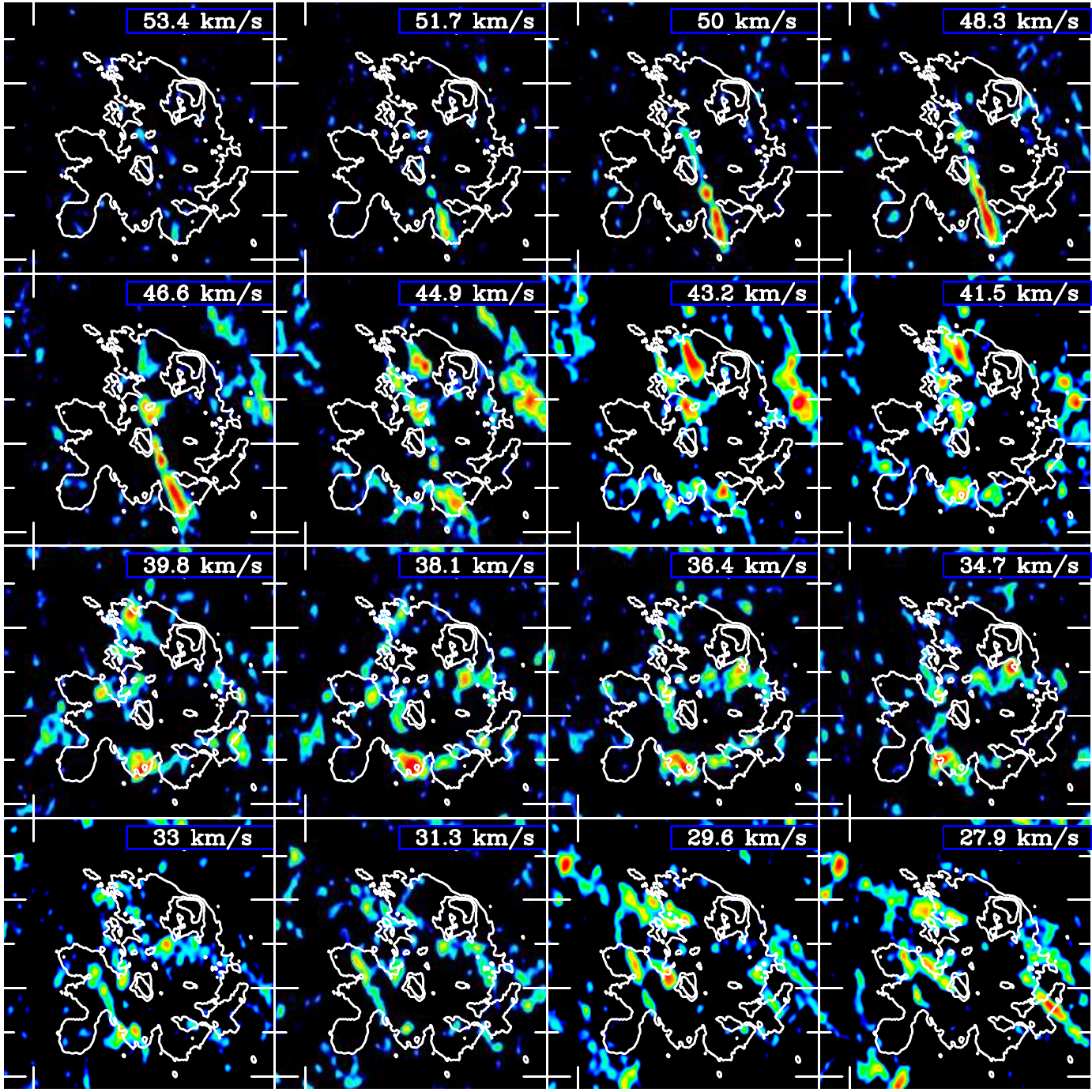}
\caption{\hi~21 cm line channel maps of G351.7 in the velocity range V=[$+27.9$~km~s$^{-1}$, $+53.4$~km~s$^{-1}$] overlaid with 321 MHz radio contours. The contour levels are 15, 100 and 200~mJy\,beam$^{-1}$. The beam is $48.6\arcsec\times32.7\arcsec$.}
\label{vel_high}
\end{figure*}

\begin{figure}
\centering
\includegraphics[scale=0.31]{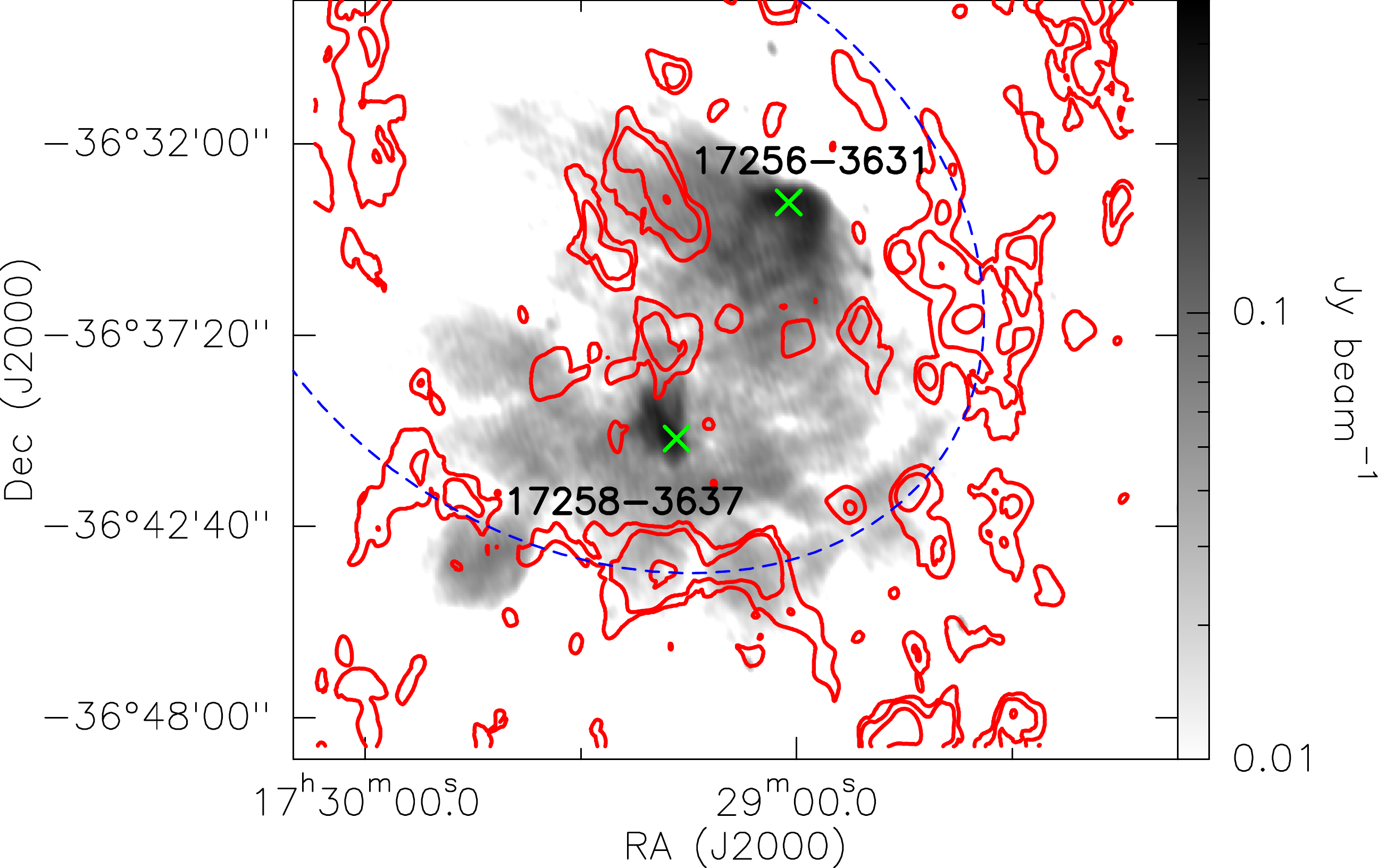}
\caption{321~MHz radio continuum emission overlaid with \hi~integrated intensity contours in the velocity range V=[$+37.9$~km~s$^{-1}$, $+44.5$~km~s$^{-1}$]. The contour levels are 22, 50 and 100~Jy\,beam$^{-1}$*m\,s$^{-1}$. Dashed arc shows the region of enhanced \hi~emission.}
\label{rgb}
\end{figure}

In order to study the distribution of \hi~emission towards G351.7 in detail, velocity channel maps have been created. The channel maps of the high velocity feature in the velocity range 27.9~km~s$^{-1}$ to 53.4 km~s$^{-1}$ are shown in Fig.~\ref{vel_high}. Most of the high velocity \hi~emission features are distributed close to the periphery of the radio shell. This is evident from Fig.~\ref{rgb}. The high velocity \hi~shell-like emission has likely origins in the expansion of atomic gas due to the supernova blast wave.

\subsubsection{Linear high velocity \hi~emission}

A closer inspection of the channel maps reveal an additional high velocity feature in the velocity range $+40$~km~s$^{-1}$ to $+52$~km~s$^{-1}$. This feature shows a linear, \hi~jet-like emission and is located towards the projected interior of the radio shell. No absorption features are observed in this velocity range. The feature is present in all the 6 individual datasets and in both RR and LL polarisations. The jet-like feature has an angular extent of $11.5\arcmin\times2\arcmin$ and is oriented along the NE-SW direction with a position angle of 19.8$^\circ$ east of north. The \hi~feature has a length of $\sim7$~pc at the assumed distance to the SNR candidate of 2~kpc, with an average width of $\sim$1~pc. The opening angle of the jet is calculated using the expression \citep[e.g.,][]{2009A&A...507L..33P} 

\begin{equation}
\alpha=2\,\taninv\left[\frac{0.5\,\sqrt{d^2-b^2}}{r}\right]
\end{equation}

\noindent where $r$ is the distance to the core along the jet axis, $d$ is the average width of the jet, $b$ is the beam size and the quantity $\sqrt{d^2-b^2}$ is the deconvolved transverse size of the jet. We find the average opening angle of the jet to be 14.4$^\circ$. This implies that the emission along the \hi~feature is highly collimated. We have not been able to identify any mid-infrared or submillimeter counterpart of this unusual feature. A Fermi-LAT $\gamma$~ray point source 1FGLJ1729.1--3641c has been found in this region. Fig.~\ref{jet_col} shows the velocity structure of the \hi~jet.

Assuming that the \hi~is optically thin, the total integrated line flux from the high velocity jet-like emission can be used to estimate the total \hi~mass ($M_{\hi}$) using the expression \citep[e.g.][]{1983ApJ...266..701K}

\begin{equation}
M_{\hi} = 0.236\,D^2 \int F dv~M_\odot
\end{equation}

\noindent where $\int F\,dv$ is the total integrated flux in units of Jy\,km~s$^{-1}$ and D is the distance in kpc. Integrating the emission within the 3$\sigma$ contour of the zeroth moment map, we find the total integrated flux to be 1.22~Jy\,km~s$^{-1}$. For a distance of 2~kpc, the total \hi~mass of high velocity feature is estimated to be 1.2~M$_\odot$.  

\subsubsection{Origin of jet-like \hi~emission}

To understand the mechanisms responsible for high velocity jet-like \hi~emission, we iinvestigated this region in other wavebands. The most common candidate for \hi~jets/outflow emission is a protostellar object. Our multiwavelength search did not reveal any such objects along the jet direction. If the emission of high velocity \hi~is due to star formation, tracers such as enhanced infrared emission and presence of submillimetre dust clumps are expected to be present. However, we do not observe any star formation tracers towards the peak of the \hi~emission or along the jet-like structure. 

\begin{figure}
\centering
\hspace*{-0.5cm}\includegraphics[scale=0.37]{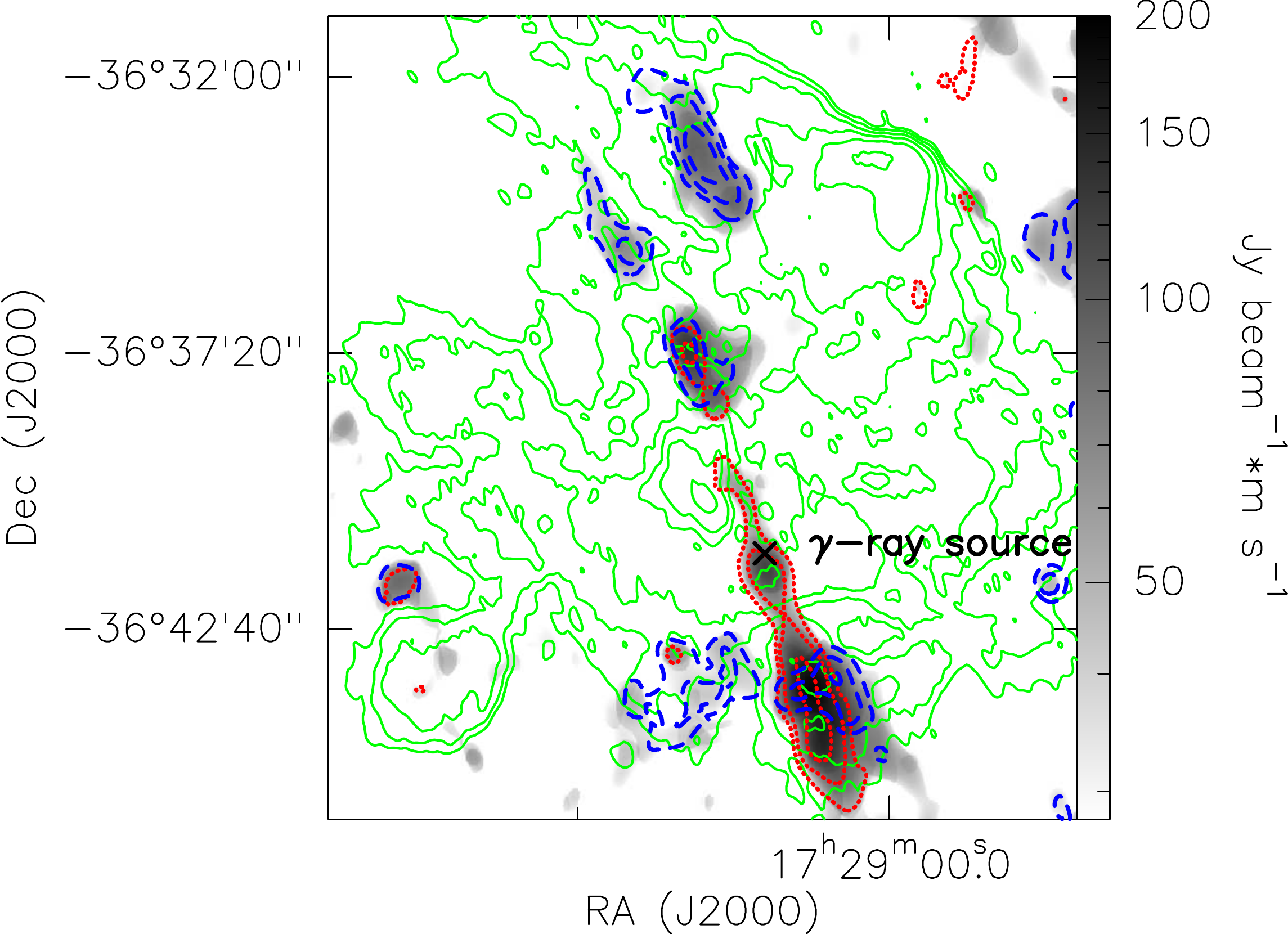}
\caption{Kinematic structure of the high velocity \hi~jet. Integrated intensity map of the high velocity jet-like feature overlaid with 321~MHz radio contours (green). Red (dotted) and blue (dashed) contours corresponds to the \hi~integrated intensity in the velocity range V$_\textrm{red}$=[46.5~km~s$^{-1}$, 51.7~km~s$^{-1}$] and V$_\textrm{blue}$=[39.7~km~s$^{-1}$, 44.9~km~s$^{-1}$], respectively. The location of the Fermi LAT $\gamma$-ray source is also marked.}
\label{jet_col}
\end{figure}

The peak of the Fermi LAT $\gamma$-ray source coincides with the peak of the high velocity \hi~emission (see Fig.~\ref{jet_col}). In Paper I, we proposed that the $\gamma$-ray source has likely origin at the site where the supernova remnant interacts with the ambient molecular cloud. Even though the angular resolution of the $\gamma$-ray source ($\sim$0.1$^\circ$) is insufficient to pinpoint the exact location, it is likely that both the high velocity \hi~emission and Fermi-LAT source are associated with each other. The association of supernova remnant candidate G351.7 and the detection of $\gamma$-ray source towards the \hi~peak indicates that the high velocity jet-like \hi~emission could have possible origins in the SNR and its progenitor. For example, \citet{1977ApJ...212..416D} report the presence of a \hi~jet associated with the Galactic SNR IC443. The evidences of interactions between relativistic jets from compact objects (such as neutron stars and black holes) and the ambient material have been reported towards few Galactic sources \citep[e.g.,][]{{1998MNRAS.299..812G},{1999ARA&A..37..409M},{2009PASJ...61L..23F}}. \citet{2002AJ....123..337D} identified a linear \hi~filament towards the SNR 320.4--1.2 which is strikingly aligned with the projected axis inferred for the collimated outflows from its central pulsar PSR B1509--58. Keeping this in view, we carried out a targeted pulsar search to identify pulsar candidate(s) associated with the SNR candidate G351.7. From our pulsar observations, we could not find any candidate pulsar associated with the region. For the on-source time of 2.8 hours with the GMRT incoherent array of 25 antennas and considering a system temperature of 160 K, we estimate the 10$\sigma$ non-detection limit as 0.18 mJy at 650 MHz over a usable bandwidth of 180 MHz for a pulsar with 10$\%$ duty-cycle.

Recently, many X-ray missions have revealed a new class of X-ray sources with no radio counterparts and very high X-ray to optical luminosity. These objects are located towards the interior of the SNRs. According to \citet{2002ASPC..271..247P}, sources showing pulsations with periods between 6 and 12s are called anomalous X-ray pulsars (AXPs) or soft gamma-ray repeaters (SGRs) if there are detections of $\gamma$-ray bursts. The remaining sources are called central compact objects (CCOs), if located towards the interior of SNR and isolated neutron stars (INSs) if there is no such association. CCOs are believed to be young, radio-quiet neutron stars. The nature and emission mechanisms of these exotic objects are still a subject of speculation \citep{2017JPhCS.932a2006D}. A double-lobed \hi~depression feature centred on the CCO of the SNR Puppis A indicates the association of \hi~jet with the CCO \citep{2003MNRAS.345..671R}. \citet{2004PASA...21...82R} propose that this double-lobed \hi~structure is created by the ejection of two opposite jets from CCO.  For our case of SNR G351.7, it is possible that the central object is a CCO with no radio signature similar to Puppis A. The \hi~jet could then arise from the bulk sweeping up of atomic gas to the boundary of the possible jet/outflow from the CCO. In order to confirm this, high resolution X-ray observations are required to characterise the nature of the central object and the origin of jet-like \hi~feature in G351.7.

\section{Conclusion}
The 21~cm spectral line and continuum observations of the supernova remnant candidate G351.7--1.2 are presented here. Faint diffuse emission from part of the low frequency radio shell is observed in the 21~cm continuum emission. \hi~absorption against the continuum is observed towards the \hii~regions IRAS 17256--3631 and IRAS 17258--3637 peaking at V$\sim-12$~km~s$^{-1}$. We have identified the high velocity \hi~emission (V$>+35$~km~s$^{-1}$) along the outer periphery of the shell. A high velocity narrow jet-like emission feature is also detected towards the interior of the radio shell in the velocity range $+40$~km~s$^{-1}$ to $+52$~km~s$^{-1}$. The mass of this jet-like feature is estimated to be 1.2~M$_\odot$. Based on the location of this high velocity feature and the presence of a Fermi-LAT $\gamma$-ray source 1FGL1729.1-3641c towards central peak of the jet-like emission, we propose a likely association with the supernova remnant and/or its progenitor. As HI jets from SNRs are relatively rare, the detection of this highly collimated jet opens up the possibility of examining them in detail to understand their origin with relation to the compact object.

\bigskip
\noindent \textbf{ACKNOWLEDGEMENT}\\

\par We thank the referee for the comments and suggestions that improved the quality of this paper. We thank the staff of GMRT, who made the radio observations possible. GMRT is run by the National Centre for Radio Astrophysics of the Tata Institute of Fundamental Research. NR acknowledges support from the Infosys Foundation through the Infosys Young Investigator grant. VVS acknowledges NR's Infosys Young Investigator grant for supporting collaborative visit to IISc to work on this project. We thank Peter Schilke for the fruitful discussions and suggestions that improved the quality of this paper.

\bibliography{ref}

\begin{thebibliography}{26}
\expandafter\ifx\csname natexlab\endcsname\relax\def\natexlab#1{#1}\fi

\bibitem[{{Cordes} \& {Lazio}(2002)}]{2002astro.ph..7156C}
{Cordes} J.~M., {Lazio} T.~J.~W., 2002, arXiv Astrophysics e-prints

\bibitem[{{De Luca}(2017)}]{2017JPhCS.932a2006D}
{De Luca} A., 2017, in Journal of Physics Conference Series, Vol. 932, Journal
  of Physics Conference Series, p. 012006

\bibitem[{{Denoyer}(1977)}]{1977ApJ...212..416D}
{Denoyer} L.~K., 1977, \apj, 212, 416

\bibitem[{{Dubner} {et~al}\mbox{.}(2002){Dubner}, {Gaensler}, {Giacani},
  {Goss}, \& {Green}}]{2002AJ....123..337D}
{Dubner} G.~M., {Gaensler} B.~M., {Giacani} E.~B., {Goss} W.~M., {Green} A.~J.,
  2002, \aj, 123, 337

\bibitem[{{Fich}, {Blitz} \& {Stark}(1989){Fich}, {Blitz}, \&
  {Stark}}]{1989ApJ...342..272F}
{Fich} M., {Blitz} L., {Stark} A.~A., 1989, \apj, 342, 272

\bibitem[{{Fukui} {et~al}\mbox{.}(2009){Fukui}, {Furukawa}, {Dame}, {Dawson},
  {Yamamoto}, {Rowell}, {Aharonian}, {Hofmann}, {de O{\~n}a Wilhelmi},
  {Minamidani}, {Kawamura}, {Mizuno}, {Onishi}, {Mizuno}, \&
  {Nagataki}}]{2009PASJ...61L..23F}
{Fukui} Y. {et~al.}, 2009, \pasj, 61, L23

\bibitem[{{Gaensler}, {Green} \& {Manchester}(1998){Gaensler}, {Green}, \&
  {Manchester}}]{1998MNRAS.299..812G}
{Gaensler} B.~M., {Green} A.~J., {Manchester} R.~N., 1998, \mnras, 299, 812

\bibitem[{Gupta {et~al}\mbox{.}(2017)Gupta, Ajithkumar, Kale, Nayak,
  Sabhapathy, Sureshkumar, Swami, Chengalur, Ghosh, Ishwara-Chandra,
  {et~al.}}]{gupta2017upgraded}
Gupta Y. {et~al.}, 2017, Current Science, 113, 707

\bibitem[{{Guzm{\'a}n} {et~al}\mbox{.}(2011){Guzm{\'a}n}, {May}, {Alvarez}, \&
  {Maeda}}]{2011A&A...525A.138G}
{Guzm{\'a}n} A.~E., {May} J., {Alvarez} H., {Maeda} K., 2011, \aap, 525, A138

\bibitem[{{Haslam} {et~al}\mbox{.}(1982){Haslam}, {Salter}, {Stoffel}, \&
  {Wilson}}]{1982A&AS...47....1H}
{Haslam} C.~G.~T., {Salter} C.~J., {Stoffel} H., {Wilson} W.~E., 1982, \aaps,
  47, 1

\bibitem[{{Herbst} \& {Assousa}(1977)}]{1977ApJ...217..473H}
{Herbst} W., {Assousa} G.~E., 1977, \apj, 217, 473

\bibitem[{{Hillebrandt}, {Nomoto} \& {Wolff}(1984){Hillebrandt}, {Nomoto}, \&
  {Wolff}}]{1984A&A...133..175H}
{Hillebrandt} W., {Nomoto} K., {Wolff} R.~G., 1984, \aap, 133, 175

\bibitem[{{Knapp} \& {Bowers}(1983)}]{1983ApJ...266..701K}
{Knapp} G.~R., {Bowers} P.~F., 1983, \apj, 266, 701

\bibitem[{{McClure-Griffiths} {et~al}\mbox{.}(2009){McClure-Griffiths},
  {Pisano}, {Calabretta}, {Ford}, {Lockman}, {Staveley-Smith}, {Kalberla},
  {Bailin}, {Dedes}, {Janowiecki}, {Gibson}, {Murphy}, {Nakanishi}, \&
  {Newton-McGee}}]{2009ApJS..181..398M}
{McClure-Griffiths} N.~M. {et~al.}, 2009, \apjs, 181, 398

\bibitem[{{Mirabel} \& {Rodr{\'{\i}}guez}(1999)}]{1999ARA&A..37..409M}
{Mirabel} I.~F., {Rodr{\'{\i}}guez} L.~F., 1999, \araa, 37, 409

\bibitem[{{Pavlov} {et~al}\mbox{.}(2002){Pavlov}, {Sanwal}, {Garmire}, \&
  {Zavlin}}]{2002ASPC..271..247P}
{Pavlov} G.~G., {Sanwal} D., {Garmire} G.~P., {Zavlin} V.~E., 2002, in
  Astronomical Society of the Pacific Conference Series, Vol. 271, Neutron
  Stars in Supernova Remnants, {Slane} P.~O., {Gaensler} B.~M., eds., p. 247

\bibitem[{{Phillips}, {Ramos-Larios} \& {Perez-Grana}(2009){Phillips},
  {Ramos-Larios}, \& {Perez-Grana}}]{2009MNRAS.397.1215P}
{Phillips} J.~P., {Ramos-Larios} G., {Perez-Grana} J.~A., 2009, \mnras, 397,
  1215

\bibitem[{{Pushkarev} {et~al}\mbox{.}(2009){Pushkarev}, {Kovalev}, {Lister}, \&
  {Savolainen}}]{2009A&A...507L..33P}
{Pushkarev} A.~B., {Kovalev} Y.~Y., {Lister} M.~L., {Savolainen} T., 2009,
  \aap, 507, L33

\bibitem[{{Reynoso} {et~al}\mbox{.}(2003){Reynoso}, {Green}, {Johnston},
  {Dubner}, {Giacani}, \& {Goss}}]{2003MNRAS.345..671R}
{Reynoso} E.~M., {Green} A.~J., {Johnston} S., {Dubner} G.~M., {Giacani} E.~B.,
  {Goss} W.~M., 2003, \mnras, 345, 671

\bibitem[{{Reynoso} {et~al}\mbox{.}(2004){Reynoso}, {Green}, {Johnston},
  {Goss}, {Dubner}, \& {Giacani}}]{2004PASA...21...82R}
{Reynoso} E.~M., {Green} A.~J., {Johnston} S., {Goss} W.~M., {Dubner} G.~M.,
  {Giacani} E.~B., 2004, \pasa, 21, 82

\bibitem[{{Roger} {et~al}\mbox{.}(1999){Roger}, {Costain}, {Landecker}, \&
  {Swerdlyk}}]{1999A&AS..137....7R}
{Roger} R.~S., {Costain} C.~H., {Landecker} T.~L., {Swerdlyk} C.~M., 1999,
  \aaps, 137, 7

\bibitem[{{Swarup} {et~al}\mbox{.}(1991){Swarup}, {Ananthakrishnan}, {Kapahi},
  {Rao}, {Subrahmanya}, \& {Kulkarni}}]{1991CuSc...60...95S}
{Swarup} G., {Ananthakrishnan} S., {Kapahi} V.~K., {Rao} A.~P., {Subrahmanya}
  C.~R., {Kulkarni} V.~K., 1991, Current Science, Vol.~60, NO.2/JAN25, P.~95,
  1991, 60, 95

\bibitem[{{Veena} {et~al}\mbox{.}(2019){Veena}, {Vig}, {Sebastian}, {Lal},
  {Tej}, \& {Ghosh}}]{2019MNRAS.482.4630V}
{Veena} V.~S., {Vig} S., {Sebastian} B., {Lal} D.~V., {Tej} A., {Ghosh} S.~K.,
  2019, \mnras, 482, 4630

\bibitem[{{Veena} {et~al}\mbox{.}(2017){Veena}, {Vig}, {Tej}, {Kantharia}, \&
  {Ghosh}}]{2017MNRAS.465.4219V}
{Veena} V.~S., {Vig} S., {Tej} A., {Kantharia} N.~G., {Ghosh} S.~K., 2017,
  \mnras, 465, 4219

\bibitem[{{Veena} {et~al}\mbox{.}(2016){Veena}, {Vig}, {Tej}, {Varricatt},
  {Ghosh}, {Chandrasekhar}, \& {Ashok}}]{2016MNRAS.456.2425V}
{Veena} V.~S., {Vig} S., {Tej} A., {Varricatt} W.~P., {Ghosh} S.~K.,
  {Chandrasekhar} T., {Ashok} N.~M., 2016, \mnras, 456, 2425

\bibitem[{{Vig} {et~al}\mbox{.}(2014){Vig}, {Ghosh}, {Ojha}, {Verma}, \&
  {Tamura}}]{2014MNRAS.440.3078V}
{Vig} S., {Ghosh} S.~K., {Ojha} D.~K., {Verma} R.~P., {Tamura} M., 2014,
  \mnras, 440, 3078

\end{thebibliography}

\end{document}